# Plasma cleaning: A new possible treatment for niobium superconducting cavity after nitrogen doping

Ziqin Yang[1], Xiangyang Lu[1)], Datao Xie[1], Lin Lin[1], Kui Zhou[2], Jifei Zhao[1], Deyu Yang[1], Weiwei Tan[1]

[1]*State Key Laboratory of Nuclear Physics and Technology, Peking University, Beijing 100871, China*
[2]*Institute of Applied Electronics, Chinese Academy of Engineering Physics, Mianyang 621900, China*

ABSTRACT

Nitrogen doping treatment with the subsequent electropolishing (EP) of the niobium superconducting cavity can significantly increase the cavity's quality factor up to a factor of 3. But the process of the EP removal may reintroduce hydrogen in the cavity surface, which may influence the cavity's radio-frequency performance. Plasma cleaning study on niobium samples with gas mixtures of argon and oxygen, intended to remove contaminations (hydrocarbons and micronicdust particles) from cavity surface to avoid field emission, was performed in Peking University. The niobium samples have been analyzed using the time of flight secondary ion mass spectrometry (TOF-SIMS) to measure the depth profiles of H, C, O, F, P and Nb. The measuring results show that the plasma cleaning with gas mixtures of argon and oxygen and conditions of about 20Pa and 100W can remarkably reduce the contents of impurity elements in the depth of about 30 nm without introducing hydrogen in the cavity surface. So plasma cleaning has been proposed to be a new possible treatment for niobium superconducting cavity after nitrogen doping.

## 1. Introduction

Quality factor Q is one of the most important factors of radio-frequency niobium superconducting cavities. High quality factor can efficiently decrease the cryogenic load of superconducting cavities, and the post treatment has an important influence on the cavities' radio-frequency performance. The average unloaded $Q_0$ of linac coherent light source (LCLS-II) 9-cell 1.3GHz cavities and cryomodules were set to be exceeding $2.7\times10^{10}$ at a gradient of 16MV/m at 2K [1]. It means that the surface resistance of the cavity is less than 10nΩ, and the standard surface treatment procedures including a combination of chemical treatments like electropolishing (EP), buffered chemical polishing (BCP) and heat treatments [2] cannot meet this requirement. Grassellino [3, 4, 5] from FNAL has reported a new surface treatment, nitrogen doping, that can systematically improve the quality factor of radio-frequency niobium superconducting cavities up to a factor of about 3 compared to the standard surface treatment procedures. Nitrogen doping consists of three steps, namely heat treatment at $800^0$C for several minutes in a nitrogen atmosphere of about 25 mTorr, heat treatment at $800^0$C for several minutes in high vacuum and EP removal of cavity material to several microns. After that Cornell University [6] and JLAB [7] have successfully repeated and demonstrated the nitrogen doping experiment on both single cell and 9-cell cavities. On one hand, Cornell University and JLAB together with FNAL have been searching for the best nitrogen doping recipe [8, 9]. On the other hand, fundamental understanding of the nitrogen doping mechanism is being carried out extensively, but yet remains unclear. But more and more studies [10, 11, 12, 13] show that hydrogen content may play an important role in the remarkable increase of the quality factor of superconducting niobium cavity after nitrogen doping treatment. The nitrogen impurities can reduce the diffusion coefficient of hydrogen in nitrogen doped niobium [14], but exorbitant nitrogen content may degrade the quality factor of the superconducting niobium cavity. The cavity directly after nitrogen treatment shows the quality factor of the order of $10^7$[3], for the possible poorly superconducting nitrides on the cavity inner surface [15]. Therefore it is important to decrease the nitrogen concentration near the surface. The EP removal after nitrogen doping treatment is to achieve this purpose. But EP removal may not be the only way to reduce the nitrogen impurity concentration near the surface after nitrogen doping. Plasma treatment has been used to the superconducting niobium cavity in Peking University since 2005 [16, 17], and some useful results have been achieved. Based on the experience of plasma cleaning study in Peking University, it has been proposed to be a new possible treatment for niobium superconducting cavity to reduce nitrogen content in the surface after nitrogen doping. The feasibility study of it is under way.

*Supported by Major Research Plan of National Natural Science Foundation of China(91026001)
*Supported by National Natural Science Foundation of China(1127506)
1)E-mail: xylu@pku.edu.cn

## 2. Plasma cleaning

Plasma treatment has been considered as an effective method to eliminate surface pollutants [18] and the application of it to the superconducting niobium cavity to get better surface has been tried [19, 20]. Plasma cleaning study on niobium samples with gas mixtures of Ar and $O_2$, intended to remove contaminations (hydrocarbons and micronicdust particles) from cavity surface to avoid field emission, was performed in Peking University. The samples are circular pillars. The bottom diameter is 10mm, the top diameter is 6mm, the total thickness is 3mm, and the thickness of cylindrical convex platform is 2mm. The samples were cut from the same niobium plate. The samples were cleaned by BCP1:1:2, removing about 75μm, and then rinsed with pure water. A total of three samples have been processed and measured before nitrogen doping treatment. The three samples were labeled as L1, L2 and L3, respectively. After that the sample L1, without any other processing, was kept in vacuum environment and was used as a reference for comparison with plasma cleaned samples. After BCP treatment, both sample L2 and sample L3 have been plasma cleaned with gas mixtures of Ar and $O_2$ under different experimental conditions. For sample L2, the Argon gas flow rate is 10sccm, the Oxygen gas flow rate is 3-4sccm, the reaction pressure is 20Pa, the reaction time is 30 minutes, and the reactive power is 110W. For sample L3, the Argon gas flow rate is 24sccm, the Oxygen gas flow rate is 5-6sccm, the reaction pressure is 50Pa, the reaction time is 30 minutes, and the reactive power is 50W.

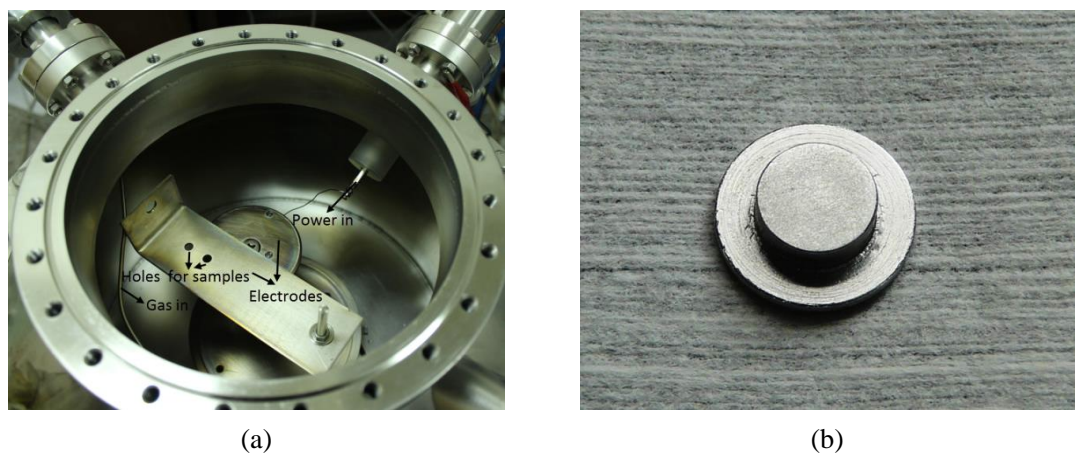

(a)          (b)

FIG. 1. The cavity (a) and sample (b) used for plasma cleaning study.

After the plasma cleaning treatment, the niobium samples have been analyzed using the time of flight secondary ion mass spectrometry (TOF-SIMS) to measure the depth profiles of H, C, O, F, P, S and Nb. $Cs^+$ primary ion beam was used to detect H, C, O, F, P, S and Nb since Cs enhances the negative ion yields. Because the surface resistance of the niobium below 2K depends largely on the penetration depth of the rf field of the order of ~40nm, it is important to get the subtle material details within $\lambda_L$. To make the depth resolution high enough, the TOF-SIMS $Cs^+$ primary ion beam energy was reduced to 2keV with the sputtering rate of 0.2nm/s. The primary ion current was 170nA; the raster area was $350\,\mu m \times 350\mu m$ with the detected area of $120\,\mu m \times 120\mu m$. In order to prevent contamination, the analysis chamber of the system was kept in the ultra-high vacuum (UHV) condition of ~$10^{-8}$Pa.

The depth profiles of H, C, O, F, P and Nb within tens of nanometers are shown in Figure 2. It can be seen that, hydrogen concentrations in the plasma treated samples are quite different from the reference. Within the depth of 10-30nm from the surface, the hydrogen concentrations in sample L2 and L3 is significantly lower than that in the reference which can reduce the possible formation of lossy hydrides.

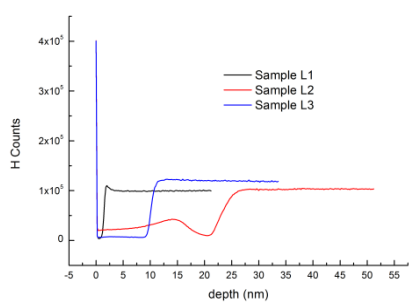

(a)

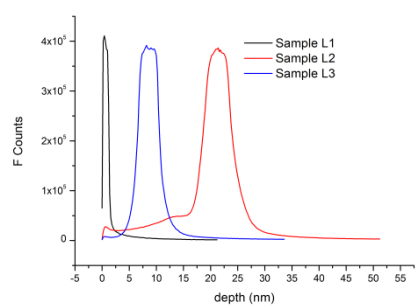

(b)

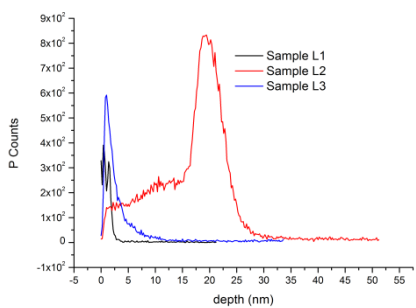

(c)

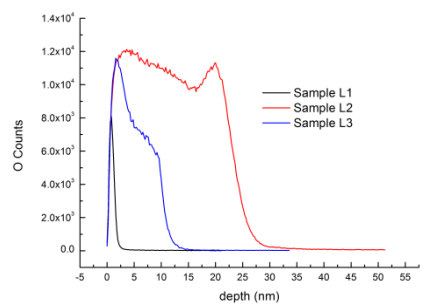

(d)

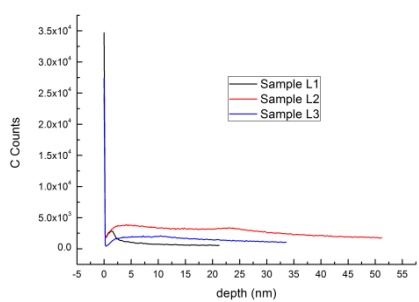

(e)

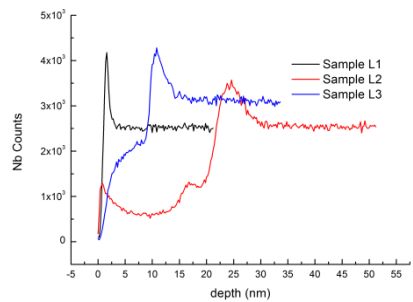

(f)

FIG. 2. The depth profiles of H (a), F (b), P (c), O (d), C (e) and Nb (f) within tens of nanometers in sample L1 (dark line), sample L2 (red line) and sample L3 (blue line). Sample L1 has not been plasma treated and is the reference, sample L2 is the heavily processed one and sample L3 is the lightly processed one.

The depth profiles of F, P, Nb show that, plasma cleaning treatment can reduce the impurity concentration near the niobium surface with the depth of tens of nanometers effectively. With the increase of reactive power and the decrease of reaction pressure, the effective depth of the plasma treatment will be larger, as can be seen in Figure 2.Because sample L2 and sample L3 were plasma cleaned with gas mixtures of Ar and $O_2$, the samples after plasma cleaning treatment have higher oxygen profiles. The effect of plasma cleaning treatment on the niobium superconductivity will be evaluated by measuring the sample's material parameters such as the energy gap, the critical temperature, the residual resistivity ratio (RRR) and the hysteresis loop.

Overall, according to the TOF-SIMS measurement, the plasma cleaning treatment can reduce the impurity concentration near the surface with the depth of about 30 nanometers, which is comparable to the penetration depth $\lambda_L$ and largely determines the niobium surface resistance. What's more, the plasma cleaning treatment does not introduce hydrogen to the niobium. So if plasma cleaning treatment is applied to niobium sample after nitrogen doping, the nitrogen concentration near the surface will be effectively reduced without reintroducing hydrogen to the niobium. And it may play the same role as the EP removal to avoid the formation of poorly superconducting nitrides. After the verification on niobium samples, the plasma cleaning treatment will be applied to the cavity and tested in Helium temperature. The experiment is under way in Peking University.

### 3. Conclusion

Nitrogen doping is a new surface treatment that can systematically improve the quality factor of radio-frequency niobium superconducting cavities up to a factor of about 3 compared to the standard surface treatment procedures. The niobium cavities after nitrogen doping treatment have to be EP processed to reduce the nitrogen concentrations near the surface, which may lead to the formation of poorly superconducting nitrides. But the EP process may reintroduce hydrogen to the cavity surface. Also the HF involved in EP process is very dangerous. Plasma cleaning study on niobium samples was performed in Peking University. The TOF-SIMS measurement shows that plasma cleaning treatment can significantly reduce the impurity concentration within the surface of about 30nm without introducing hydrogen into the niobium. The subtle material details within this depth largely determine the niobium cavity surface resistance. So the plasma cleaning treatment may be applied to reduce the nitrogen concentration near the surface. Study about the feasibility is under way.